\documentstyle[preprint,aps]{revtex}
\input epsfig.sty
              
\begin{document}
\title{The quantum mechanical geometric phase of a particle in a 
resonant vibrating cavity}
\author{K.~W.~Yuen, H.~T.~Fung, K.~M.~Cheng, M.~-C.~Chu, and K.~Colanero}
\address{Department of Physics, The Chinese University of Hong Kong, 
Shatin, N.T., Hong Kong.}
\maketitle
\tightenlines

\begin{abstract}
We study the general-setting quantum geometric phase acquired by a particle 
in a vibrating cavity.
Solving the two-level theory with the rotating-wave approximation 
and the SU(2) method, 
we obtain analytic formulae that give excellent descriptions
of the geometric phase, energy, and wavefunction of the resonating system.
In particular, we observe a sudden $\pi$-jump in the geometric phase when
the system is in resonance.  We found similar behaviors in the geometric
phase of a spin-1/2 particle in a rotating magnetic field, for which we
developed a geometrical model to help visualize its evolution.

\end{abstract}

\narrowtext

\section{Introduction}
The dynamics of confined cavity fields interacting with the cavity 
wall is of great interest for the understanding of a variety of problems such 
as hadron bag models \cite{bag}, sonoluminescence \cite{sono}, 
cavity QED \cite{qed} and black hole radiations \cite{bh}.
Most previous studies on dynamical cavity concentrated on scalar or 
photon fields \cite{photon}, which despite the apparent simplicity,
exhibit rich and complex dynamics. 
In contrast, the system of a Schr\"odinger particle in an oscillating 
spherical cavity has not gained as much attention. 
In this article, we study the generalized quantum geometric phase of a 
particle in resonance with the vibrating cavity. 
We will show that the geometric phase acquires sudden $\pi$-jumps when
the particle makes transitions.

The geometric phase of a quantum system has drawn much 
attention since it was first proposed by Berry in 1984 
\cite{berry,Alfred}. It reveals the geometric structure of 
the Hilbert space of the system \cite{Alfred,Hilbert}, 
and its existence in physical systems has 
been verified in many experiments \cite{moore}, including electron diffraction 
from a screw dislocation \cite{bird} and neutron interferometry \cite{wagh}. 
The phase effects in molecular physics can lead to energy splittings and shift 
quantum numbers. The geometric phase has also been shown to be intimately
connected to the physics of fractional statistics, the quantized Hall effect, 
and anomalies in gauge theory \cite{Alfred}.
As far as we know, our study represents the first calculation of the
geometric phase of a resonating system, which evolves non-adiabatically and
non-cyclically.

\section{Formalism}
We consider an infinite cylindrical or spherical potential well with 
oscillating boundary \cite{klaus}:
\begin{equation}
  V(r) = \left\{
                 \begin{array}{cc}
                    0, & \mbox{if $r < R(t)$},\\
                    \infty, & \mbox{if $r \ge R(t)$},
                 \end{array}
           \right.
\end{equation}  
where $R(t)=R_{0}/\alpha (t)$, with $\alpha (t) \equiv 
\left[1+\epsilon \sin (\omega t)\right]^{-1}$. The coordinates 
can be transformed to a fixed domain via $\vec{y} \equiv \alpha(t)\vec{r}$,
and, to preserve unitarity, the wavefunction is renormalized through
$\phi(\vec{y},t)\equiv \alpha ^{-\xi}\psi (\vec{r},t)$,
where $\xi = 1$ (3/2), for a cylindrical (spherical) well. 

Since the full Hamiltonian $H(t)$ commutes with $L^{2}$ and $\vec{L}$, 
the wavefunction can be factorized:
$\phi(\vec{y},t)=Y(y,t)\Theta$, where $\Theta$ depends only on the angular 
variables. Inside the cavity, the radial wavefunction satisfies
\begin{equation}
\label{d.r.e}   
 \frac{\partial Y}{\partial t}
=\left[\frac{i\hbar \alpha ^2 }{2\mu }
 \left(\frac{\partial^{2}}{\partial y^{2}}+
\frac{n_d}{y}\frac{\partial}{\partial y}-\frac{m_d}{y^{2}}\right)
+ \frac{\dot{R}}{R} \left( y\frac{\partial}{\partial y} + \xi \right)\right]Y,
\end{equation}
where $\mu$ is the particle mass, and $n_d = 1$, 2 and $m_d = m^2, \ l(l+1)$ 
for cylindrical and spherical
wells respectively. In this paper, we only consider the $m_d=0$ sector.

The wavefunction described by Eq.~\ref{d.r.e} evolves in time and
acquires a time-dependent phase, which in general consists of a
dynamical phase and a geometric one \cite{berry}. 
When the dynamics is adiabatic and/or the evolution is cyclic, 
the geometric phase, or 
Berry's phase, has been studied for many systems. Since we are
interested in the geometric phase of a non-adiabatic, non-cyclic system, 
we have to resort to a generalized method.
Following Ref.~\cite{gs}, we first remove the dynamical phase from
the wavefunction of the system:
$|\phi(t)\rangle \rightarrow |\tilde{\phi}(t)\rangle \equiv  
e^{-i\theta(t)}|\phi(t)\rangle \ $, the dynamical phase is
\begin{equation}
 \theta(t) \equiv -\frac{1}{\hbar}\int_{0}^{t}E(t')dt'\;,
\end{equation}
where $E(t')={\it Re}\langle\phi(t')|H(t')|\phi(t')\rangle$.

The general setting geometric phase, or the Pancharatnam phase $\beta$ is 
defined as the relative phase between the state at 
time $t_{1}$ and that at time $t$, both with the dynamical phase removed.
It can be obtained from the inner product of these two states
\begin{equation}
\beta (t_1, t)\equiv -i \log \left[  \langle \tilde{\phi}(t_{1})|
\tilde{\phi}(t)\rangle \over | \langle \tilde{\phi}(t_{1})
|\tilde{\phi}(t)\rangle | \right] .
\label{beta}
\end{equation}
Our main goal in this paper is to study this geometric phase for a
simple but nontrivial dynamical system.

\section{Results}
We have solved Eq.~\ref{d.r.e} numerically 
and checked that the solution is stable and
converges very well.  In Fig.~1, we show the maximum energy of the particle
as a function of the driving frequency $\omega$, both having been scaled 
to be dimensionless: $\tilde{E}_{\rm max} \equiv \mu 
R_0^2E_{\rm max}/\hbar^2$, and $\tilde{\omega}\equiv \mu R_0^2 \omega/\hbar$. 
The particle is initially in the ground state, and it is in resonance at
specific values of $\tilde{\omega}$.  In each period of cavity vibration  
$\tau \equiv 2\pi/\tilde{\omega}$, the geometric phase acquires some changes,
$\beta _1 (t_1) \equiv \beta(t_1,t_1+\tau)$, and it
exhibits oscillations at these resonances. We show three
examples in Fig.~2 for $\tilde{\omega}=66.632,\ 17.278$ and 22.21227,
where the geometric phase acquires periodic changes of $2\pi, \ 4\pi, \ 6\pi$
respectively. All the resonances shown in Fig.~1
are associated with oscillations of $\beta _1$ with amplitudes of $2N\pi$,
$N = 1,\ 2,\ 3$. 

To understand the resonances and the associated geometric phases, we limit
ourselves to the parameter regime where the Hamiltonian
can be truncated to a two-level system.  Specifically, if  
$\epsilon <<1$, and 
the initial state is the $k^{\rm th}$ unperturbed eigenstate 
$|\phi_{k}(y)\rangle$, with eigenenergy $E_{k}$, then when the driving 
frequency corresponds to 
the energy difference between the initial state and another unperturbed 
eigenstate $|\phi_{n}(y)\rangle$, with eigenenergy $E_{n}$, {\it i.e.}, 
$\omega=\omega_{nk} \equiv (E_{n}-E_{k})/\hbar$,
the particle is expected to behave as in a two-level system. 
Then the problem simplifies considerably, and we 
have obtained its solution with two analytic approaches:
the SU(2) method and the Rotating-wave Approximation (RWA).

Following Cheng {\it et al.}\cite{c.m.cheng} 
we first expand the time-dependent Hamiltonian $H(t)$ of a two-level system
in the identity operator $I$, the raising and lowering operators 
$\sigma _{\pm}$, and Pauli spin matrix $\sigma_{3}$
\begin{equation}
H(t)=i \left[f_0(t)I +  f_1 (t) \sigma_+ + f_2 (t) \sigma_3 
          + f_3 (t) \sigma _- \right] , 
  \label{Eq.H(t)}
\end{equation}
where $f_j, \ j=0,1,2,3$, are in general complex functions of time.
The evolution operator can then be written as
\begin{eqnarray}
\label{U}
  U(t,0) 
=
  b(t)e^{\int^t_0 f_0(t')dt'}
  \left( \begin{array}{cc} 
	    b^{-2}(t)+g_{1}(t)g_{3}(t) & g_{1}(t) \\
	    g_{3}(t)             & 1 
	 \end{array}
  \right) ,
\end{eqnarray}
where $b(t) \equiv \exp(-g_2(t))$, and $g_i$ satisfies
\begin{eqnarray}
 \dot{g_1} &=& f_1 + 2f_2g_1-f_3g_1^2 \ , \\
 \dot{g_{2}} &=& f_{2}-f_{3}g_{1} \ , \\
\label{g3dot}
 \dot{g_{3}} &=& f_{3}b^{-2} \ ,
\end{eqnarray}
with the initial conditions $g_j(0)=0$ for $j=$1, 2, or 3.

Suppose that the two levels in which the system oscillates are 
$|\phi_{k}(y)\rangle$ and $|\phi_{n}(y)\rangle$, so that the 
wavefunction is $ |\phi(y,t)\rangle=a_{k}(t)|\phi_{k}(y)\rangle
+a_{n}(t) |\phi_{n}(y)\rangle$. Furthermore, if the initial conditions are 
$a_{k}(0) = 1$ and $a_{n}(0) = 0$, then
\begin{equation}
  \left( \begin{array}{c}
            a_k(t) \\
            a_n(t)
         \end{array}
  \right)
=
  b(t)e^{\int^t_0 f_0(t')dt'}
  \left( \begin{array}{c} 
	    b^{-2}+g_{1}g_{3} \\
	    g_{3}                  
	 \end{array}
  \right)  .
\end{equation}
The SU(2) method is exact and reduces the problem to solving the ODE's
for $g_j$.  

The spirit of the RWA is to retain
only those terms in the Hamiltonian that correspond to the resonance
frequency.
We first separate out the fast phase factors from the wavefunction:
\begin{equation}
\label{2.level.phi}
|\phi(y,t)\rangle=
c_{k}(t)e^{-i\rho_{k}(t)}|\phi_{k}(y)\rangle+
c_{n}(t)e^{-i\rho_{n}(t)}|\phi_{n}(y)\rangle,
\end{equation}
with
$\rho_{i}(t) \equiv E_{i}\int_{0}^{t}\alpha^{2}(t')dt'/\hbar$.
Substituting Eq.~\ref{2.level.phi} in the Schr\"odinger equation, we have
\begin{eqnarray}
\label{matrix.eq.b}
    \left( \begin{array}{c}
	         \dot{c_{k}} \\
	         \dot{c_{n}}
	       \end{array}
	\right)
=  \eta _{nk} \left( \begin{array}{cc}
	         0 & -W^{*}(t) \\
	         W(t) & 0
	       \end{array}
    \right)
    \left( \begin{array}{c}
	         c_{k} \\
	         c_{n}
	       \end{array}
    \right),
\end{eqnarray}
where
\begin{equation}
 W(t) \equiv \frac{\dot{R}(t)}{R(t)}
    \exp \left[ i\omega_{nk}\int_{0}^{t}\alpha^{2}(t')dt' \right],
\end{equation}
and
\begin{equation}
  \eta_{nk} 
\equiv 
  \left\langle\phi_{n}\mid\left( y\frac{\partial}{\partial y} +
  \xi \right)\mid\phi_{k} \right\rangle
\end{equation}
is a constant that depends only on the states involved. 

As $\epsilon$ is small, we can expand $W(t)$ as a series of $\epsilon$.
Retaining all terms up to third-order, we have:
\begin{eqnarray}
 W(t) &=&  \epsilon e^{i\omega_{nk}t} \left\{ \omega\cos \omega t  
  +\epsilon  \left[ i \omega_{nk}(\cos 2\omega t +1)-
   {\omega\over 2}\sin 2\omega t \right]  \right. \nonumber \\
 &+ &  {\epsilon^{2} \over 4\omega}
    \left[ \left(\omega ^2-6\omega_{nk}^2 \right)
\cos \omega t - \left(\omega^2+2\omega_{nk}^2\right) \right.
\cos 3\omega t \nonumber \\ 
   &-&  \left. \left. 
i\frac{7}{2} \omega_{nk}\omega(\sin \omega t + \sin 3\omega t) 
\right] \right\} .
\label{W}
\end{eqnarray}

Note that $W(t)$ consists of oscillatory terms with various frequencies
depending on $\omega$.
In the spirit of RWA, we keep only terms with the {\it lowest} frequency
for each $\omega$, the rationale being that $c_k$ and $c_n$ vary slowly
in time and the contributions to $W$ from high frequency terms cancel
on average over such a long time scale. It is clear then from Eq.~\ref{W} 
that $W$ is largest if 
$\omega = \omega_{nk}/N$, with $N$ an integer, because of the emergence of
zero frequency terms.  For these driving frequencies, $W$ is large and 
effective in inducing transitions, and we have resonances.
At or close to a resonance, Eq.~\ref{W} simplifies tremendously, and 
closed-form solutions for Eq.~\ref{matrix.eq.b} can be obtained
easily.

For example, when 
$\delta \omega \equiv \omega- \omega_{nk} \approx 0$, 
$W(t) \approx \epsilon \omega \exp \left(-i \delta \omega t/2 \right)/2$,
and we have Rabi oscillations \cite{sakurai}:
\begin{eqnarray}
\label{c_k}
  c_{k}(t)
&=& e^{i\delta \omega t/2}
  \left[ \cos \chi t - 
         i\delta\omega\sin \chi t/ (2\chi)  \right] , \\
\label{c_n}
  c_{n}(t)
&=& \Gamma  e^{-i\delta \omega t/2}\sin \chi t /\chi , 
\end{eqnarray}
where  we took $c_k(0)=1, \ c_n(0)=0$, 
$\chi \equiv \sqrt{  \Gamma^{2} +\delta \omega^{2}/4 }$, and  
$\Gamma = \Gamma _1 \equiv \eta _{nk} \epsilon \omega /2$.
The maximum value of $|c_{n}|^{2}$,  
$|c_{n}|_{\rm max}^{2}(\omega)=1/[1+(\delta \omega /2 \Gamma)^2]$,
is a Lorentzian with a FWHM of $4\Gamma$.

The solutions for higher $N$ resonances differ only in the 
widths $\Gamma = \Gamma _N \equiv \epsilon^N\eta_{nk}\gamma_{N}/2N $, where
\begin{equation}
{\gamma _N} = \left\{ \begin{array}{cc}
                                (\omega _{nk}+\delta \omega) & N=1, \\
     \left( 3\omega_{nk} - \delta \omega \right)/4  & N=2, \\
{17\omega_{nk}^{2} - 17\omega_{nk} \delta \omega 
+ 2\delta \omega^{2}\over 24(\omega _{nk}+\delta \omega)}  & N=3,
\end{array}
\right.
\end{equation}
with $\delta \omega \equiv N\omega-\omega_{nk}$.  The FWHM
of the resonances $4\Gamma/N \propto \epsilon ^N$ are narrower for larger $N$,
and the oscillation periods $T \propto 
\epsilon ^{-N}$ are longer.

The energy of the system is given in RWA by
\begin{eqnarray}
\label{approx.Et.1}
   E(t)
\approx   \alpha^{2}(t)
    \left[ E_{k}\cos^{2} \chi t + (A + E_{n})\sin^{2} \chi t 
    \right], 
\end{eqnarray}
where $A \equiv  \left[E_{k}\delta \omega ^2+
   4E_{n} \left( \Gamma ^2-\chi^{2}\right) 
	   \right]/4\chi^2$ .
The energy of the particle therefore exhibits fast but small oscillations 
due to the $\alpha^2$ factor, modulated by a large but slow oscillation 
of period $\pi / \chi$.  It follows that
\begin{eqnarray}
\label{transform.Emax}
 \frac{ (1-\epsilon)^{2}E_{\rm max} - E_{k} }{ E_{n} - E_{k} }
\approx
 \frac{ 1 }{1 +\left( \delta \omega/2\Gamma \right)^{2}} 
=|c_{n}|^{2}_{\rm max}(\omega) \ ,
\end{eqnarray}
which is again a universal Lorentzian for all resonances.

The RWA results are virtually identical with those of the
SU(2) method. Fig.~1 can be understood completely 
in terms of overlapping series of the $\omega _{nk}/N$ resonances,
and the positions of the peaks are in almost exact agreement with the
RWA or SU(2) predictions.  The resonance line shapes are well fitted by
the Lorentzian Eq.~\ref{transform.Emax}, though the widths are
underpredicted for the $N=2, ~3$ resonances, as shown in Table 1.
For example, the $(N=2, n=4)$ resonance has a much larger width than
predicted by RWA and SU(2), which is mainly due to 
the involvement of other states during transition, such as a $(1\rightarrow 3
\rightarrow 2)$ process, which is second order in perturbation theory and
so affects the $N>1$ transitions more severely. 
Occasionally, resonances at similar frequencies may overlap and lead to
broadened widths.

At a resonance, the Rabi oscillations of the particle as predicted in RWA
(Eq.~\ref{approx.Et.1}) can be seen explicitly in Fig.~2, where the
energy of the particle vs.~time is shown for $n=4,3,4$ and $N=1,2,3$. The
RWA also predicts that at exact resonances, $\chi = \Gamma$, 
and therefore the oscillation periods 
should be inversely proportional to the widths of the resonances,  
$T=\pi/\chi =\pi/\Gamma$. 
As shown in the last column of Table 1, this is
clearly borne out in the numerical data, where $T\Gamma/\pi$ is listed
and is found to be closed to 1 for all resonances, deviating by less than
10\% for even those with strong mixing.

Using the RWA and two-level approximation, we can calculate the 
dynamical phase easily.  Removing the dynamical phase with the help of
Eq.~\ref{approx.Et.1}, we have
\begin{equation}
        |\tilde{\phi}(t)\rangle \approx 
e^{i\omega'_{nk}
	  \int_{0}^{t}\sin^{2}(\chi t')dt' }
   \left(   \begin{array}{c}
       c_{k}(t) \\
       c_{n}(t)e^{-i\omega_{nk}t}
   \end{array} \right), 
\end{equation}
where $\omega'_{nk} \equiv  A/\hbar + \omega_{nk}$.  
For $\omega = \omega_{nk}/N$ resonances,
if we choose $t - t_1$ to be an integral multiple, $q$, of the cavity
oscillation period $\tau$, we get
\begin{equation}
 \beta (t_1, t_1+q\tau)={\omega'_{nk} \over 2\chi} 
   \left[{\tilde \chi}-\sin {\tilde \chi}
    \cos (2\chi t_1+{\tilde \chi}) \right],
\nonumber
\end{equation}
where ${\tilde \chi} \equiv \chi q \tau$.  In particular, if $t_1=0$, 
\begin{equation}
\beta_0(q\tau)\equiv\beta(0,q\tau)=
 \left\{ \begin{array}{cc} 
          \frac{\Omega(q\tau)}{2},       & {\rm for\ } (2m-{1\over 2}) < t/T 
\leq (2m+{1\over 2}), \\
          \frac{\Omega(q\tau)}{2} \pm \pi, & {\rm for\ } (2m+{1\over 2}) < 
t/T \leq (2m+{3\over 2}), 
	 \end{array}
 \right .
\label{beta0.vib}
\nonumber
\end{equation}
where
$   \Omega(t)=
 \omega_{nk}t-\frac{\omega_{nk}}{2\chi}\sin 2\chi t$ ,
with $m=0,1,2,...$. There are sudden approximate
$\pi$ jumps in $\beta_0$ at $t/T=(2m+1)/2$, 
as shown for example in Fig.~3 for $\tilde{\omega}=12.344$. 
Since $\chi \tau << 1$, the phase change in each cycle ($q=1$) is
\begin{equation}
\label{ngbp1}
\beta_1 (t_1) \equiv \beta (t_1,t_1+\tau) \approx \omega'_{nk}\tau 
\sin^2 \chi(t_1+\tau/2) .
\end{equation}
At an exact $N^{\rm th}$ resonance, $\delta \omega = 0$ and $A=0$, 
and so
\begin{equation}
\label{beta1}
   \beta_1 (t_1) \approx 
   2N\pi \sin^{2}\left[\chi (t_1+N\pi/\omega _{nk}) \right].
\end{equation}
Therefore, the geometric phase oscillates with an amplitude of $2N\pi$ and 
period of $T$.  Both of these RWA predictions are in excellent
agreement with the numerical data, as shown in Fig.~2. Note that the 
$\pi$-jumps and the functional forms of the geometric phases
are independent of $\epsilon$ and $\hbar$, as long as they are nonzero.

\section{An Electron in a Magnetic Field}
To gain more insight into the geometric phase for a two-level system, 
we have studied a simple model
of a magnetic field rotating around a spin-1/2 particle.
Suppose an electron of charge $-e$ and mass $m$ is placed at the origin, in
the presence of a magnetic field
\begin{equation}
 {\bf B}(t)
= B_0[\sin\alpha\cos(\omega t)\hat{i} + \sin\alpha\sin(\omega t)\hat{j}
      + \cos\alpha\hat{k} ]\;,
\end{equation}
which has a constant magnitude $B_0$ but its direction sweeps out a cone
with an opening angle $\alpha$, $0< \alpha < \pi$, at a constant
angular speed $\omega$.  The Hamiltonian of the system is given by
\begin{equation}
\label{ham}
 H(t)
= {e \over m}{\bf B}\cdot {\bf S}
= -{\hbar \omega_1 \over 2} \left( \begin{array}{cc}
                                 \cos\alpha & e^{-i\omega t}\sin\alpha \\
                                 e^{i\omega t}\sin\alpha & -\cos\alpha
                                 \end{array}
                            \right)\;,
\end{equation}
where $\omega_1 \equiv -eB_0/m$ and {\bf S} is the spin matrix.

The system can be solved analytically \cite{griffiths}. 
The instantaneous eigenspinors of $H(t)$ with eigenenergies 
$E_+=-E_-=-\hbar\omega_1/2$ respectively are
\begin{equation}
\label{psi+}
 |\psi_+(t)\rangle = \left( \begin{array}{c}
                    \cos(\alpha/2) \\
                    e^{i\omega t}\sin(\alpha /2)
                    \end{array} \right)
\end{equation}
and
\begin{equation}
\label{psi-}
 |\psi_-(t)\rangle = \left( \begin{array}{c}
                    \sin(\alpha/2) \\
                    -e^{i\omega t}\cos(\alpha /2)
                    \end{array} \right)\;.
\end{equation}

Suppose the electron spin is 
initially parallel to ${\bf B}(0)$, and
we consider the case when the particle makes a transition to spin down
along the instantaneous direction of ${\bf B}$.  This happens with unit 
probability if $\omega=-\omega_{1}/\cos\alpha$, provided that $\cos \alpha
\neq 0$.  The state vector at any time is then \cite{griffiths}
\begin{eqnarray}
\label{simp.psi}
 |\psi(t)\rangle =  e^{-i\omega t/2} \left\{ \cos \left({\lambda t \over 2} 
\right) |\psi_+(t)\rangle
           +i\sin \left({\lambda t \over 2} \right)
           |\psi_-(t)\rangle \right\}\; ,
\end{eqnarray}
where 
$\lambda=\omega \sin\alpha\;.
\label{lambda}$

The Pancharatnam phases comparing the initial state with the state at time
$t$ for different values of $\alpha$ are shown in 
Fig.~(\ref{ebb0k100w60}-\ref{ebb0k100.2w90}).
The sudden $\pi$-jump and a two-period oscillation can be seen in the case 
when 
$\alpha=1/p$, $p$ being an even integer, such as the case in 
Fig.~(\ref{ebb0k100w60}). When $p$ is an odd integer,
the Pancharatnam phase performs a single-period oscillation with no 
$\pi$-jump, as shown in Fig.~(\ref{ebb0k101w61}).
It has neither single-period nor two-period oscillation when
other values of $\alpha$ are used, Fig.~(\ref{ebb0k100.2w90}).
If $t=q\tau$, with $q$ an integer and $\tau =2 \pi/\omega$, 
\begin{equation}
\beta_0(q\tau)=
 \left\{ \begin{array}{cc}
          -{\Omega(q\tau) \over 2}, & {\rm  for}\; 
(2m-\frac{1}{2})<q\tau/T\leq (2m+\frac{1}{2}), \\
          -{\Omega(q\tau) \over 2}\pm \pi, & {\rm  for}\;
(2m+\frac{1}{2})<q\tau/T\leq (2m+\frac{3}{2}),
         \end{array}
 \right .
\label{beta_0.spin}
\end{equation}
which has the same form as Eq.~\ref{beta0.vib}, where $m$ 
is an integer, $T=2\pi/\lambda$, and \begin{equation}
\Omega(t) \equiv \left({\omega_1\over \lambda}\sin{\lambda t}
                     + \omega t \right) .
\label{Omega.spin}
\end{equation}
For the case $t_1 =(q-1) \tau$ and $t= q\tau$,
\begin{equation}
\beta_1(t_1)=
\left\{
\begin{array}{cc}
          -{1\over 2} \left[ \Omega (t_1+\tau) - \Omega (t_1) \right]
\ ,& {\rm  for}\;
0<\sin\alpha<1/2,\\
-{1\over 2} \left[ \Omega (t_1+\tau) - \Omega (t_1) \right] \pm \pi\ ,
& {\rm  for}\;
1/2<\sin\alpha<1\;.  
\label{beta1.k1}
\end{array}
\right.
\end{equation}
In the limit $\alpha\ll1$,
\begin{eqnarray}
\beta_{1}(t_{1})&\approx&-2\pi\sin^2\left(\frac{\lambda}{2}
(t_{1}+\tau/2)\right)\;, 
\end{eqnarray}
which has the same form as Eq.~\ref{beta1}.

In the cyclic limit, {\it i.e.}, $t=q\tau = T \equiv 2\pi/\lambda $,
\begin{equation}
 \Omega(T) = \omega T = \omega q\tau =2 q\pi \ ,
\end{equation}
and thus the geometric phases are
$\beta_0(T) =-(q-1)\pi $ and $\beta_1(T) = -\pi $.

Since the geometric phases in the two models - an electron in a rotating
magnetic field and a particle in a vibrating cavity - 
are remarkably similar,     
we conjecture that the main features of the generalized geometric phase
we calculated, especially the $\pi$-jumps, are universal for a particle 
in transition from one state to another.
Similar features about the geometric phase have also been obtained in 
\cite{R. Bhandari}.

\section{Geometrical model} 
Here we present a geometrical model that helps to visualize the 
evolution of the geometric phase of a two-level system.
It is clear from Eq.~\ref{simp.psi} that the state vector traces out
a path on the unit sphere defined by the angular coordinates 
$\theta \equiv \lambda t$ and $\phi \equiv \omega t$.
Note that we have used a time-dependent basis  
$|\psi_+(t)\rangle$, $|\psi_-(t)\rangle$, which mark
the North and South Poles respectively on the unit sphere.
The excitation condition of our system therefore gives us
the trajectory of the state on this unit sphere, 
\begin{equation}
\theta = \phi \sin \alpha \ ,
\end{equation} 
which is a spiral curve.  

The solid angle subtended at the origin  
by the spiral curve and the geodesic, $\phi=0$, up to any $t=q \tau$ is
\begin{eqnarray}
 {\Omega}_o&=& \int_0^{2 q \pi}\int_0^{\phi'\sin\alpha}\sin\theta'd\theta'd\phi'\\
       &=& 2q\pi -{\sin{(2q\pi\sin\alpha)} \over \sin\alpha} , 
\end{eqnarray}
which coincides with Eq.~\ref{Omega.spin},
in the limit $\alpha  \ll 1$. 
Therefore, we see from Eq.~\ref{beta_0.spin} and Eq.~\ref{beta1.k1} that  
the geometric phases $\beta_0$ and $\beta_1$ are simply the solid angles 
subtended by the spirial curve, $\theta=\alpha \phi$, from $\phi = 0$
and $\phi = 2(q-1)\pi$ up to $\phi = 2q\pi$ respectively, divided by $-2$.
This picture is a generalization of Berry's \cite{berry} for 
adiabatic and cyclic evolution corresponding to trajectories with
constant $\theta = \theta _0$, and the solid angle is simply
${\Omega}_o = 2 \pi (1-\cos \theta_{0})$.
Even the $\pi$-jump can be represented
in this model by a jump of the particle from one sphere to another
if its trajectory happens to reach the South Pole, which depends on the 
value of the angle $\alpha$.   The double-sphere picture reminds one
of the underlying SU(2) structure.

\section{Summary}
In summary, we have reported the first calculation of the quantum
geometric phase of a physical system in resonance - that of a particle
in a vibrating cylindrical or spherical cavity, and we have shown that
it acquires sudden $\pi$-jumps when the particle makes transitions from
one state to another.  
We have derived analytic expressions in the RWA and SU(2) methods,
which give excellent description of the energy, wavefunctions, and the 
geometric phases at these resonances.  We found remarkably similar
properties of the geometric phases for the simple system of an
electron in a rotating magnetic field, which led us to conjecture that
the main features of the generalized geometric phases and especially 
the $\pi$-jump we found are universal.
We have also developed a geometrical model to help visualize these phases.

This work is partially supported by a Chinese University Direct Grant
(Project ID: 2060093).

\begin{figure}
\psfig{file=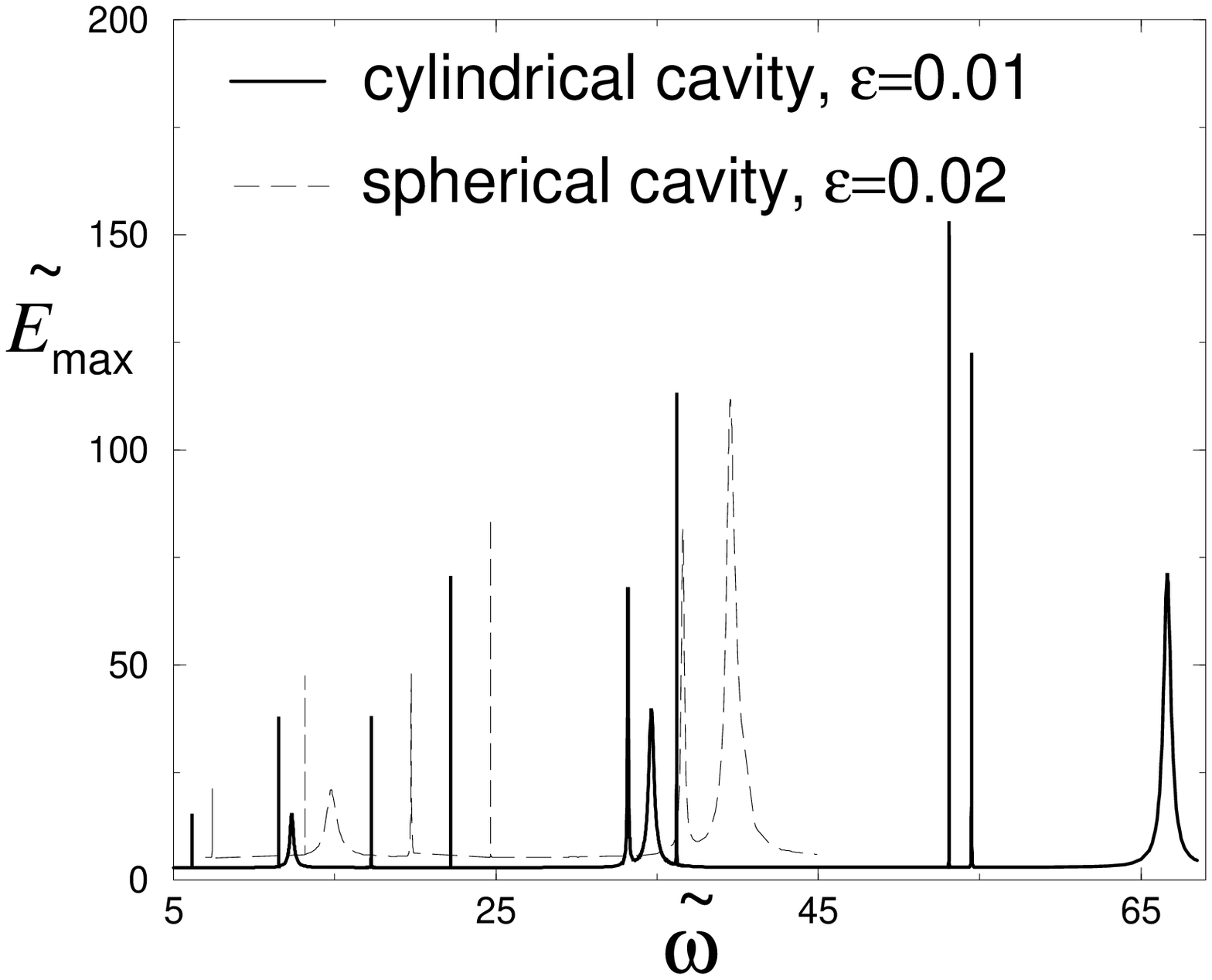,angle=0,width=12cm}
\caption{\small The maximum energy of the particle as a function of
  the driving frequency, obtained by solving the Schr\"odinger equation 
  numerically, for a cylindrical cavity with $\epsilon = 0.01$ 
  (solid line) and a spherical cavity with $\epsilon = 0.02$ (dashed line).}
\label{fig1}
\end{figure}

\begin{figure}
\psfig{file=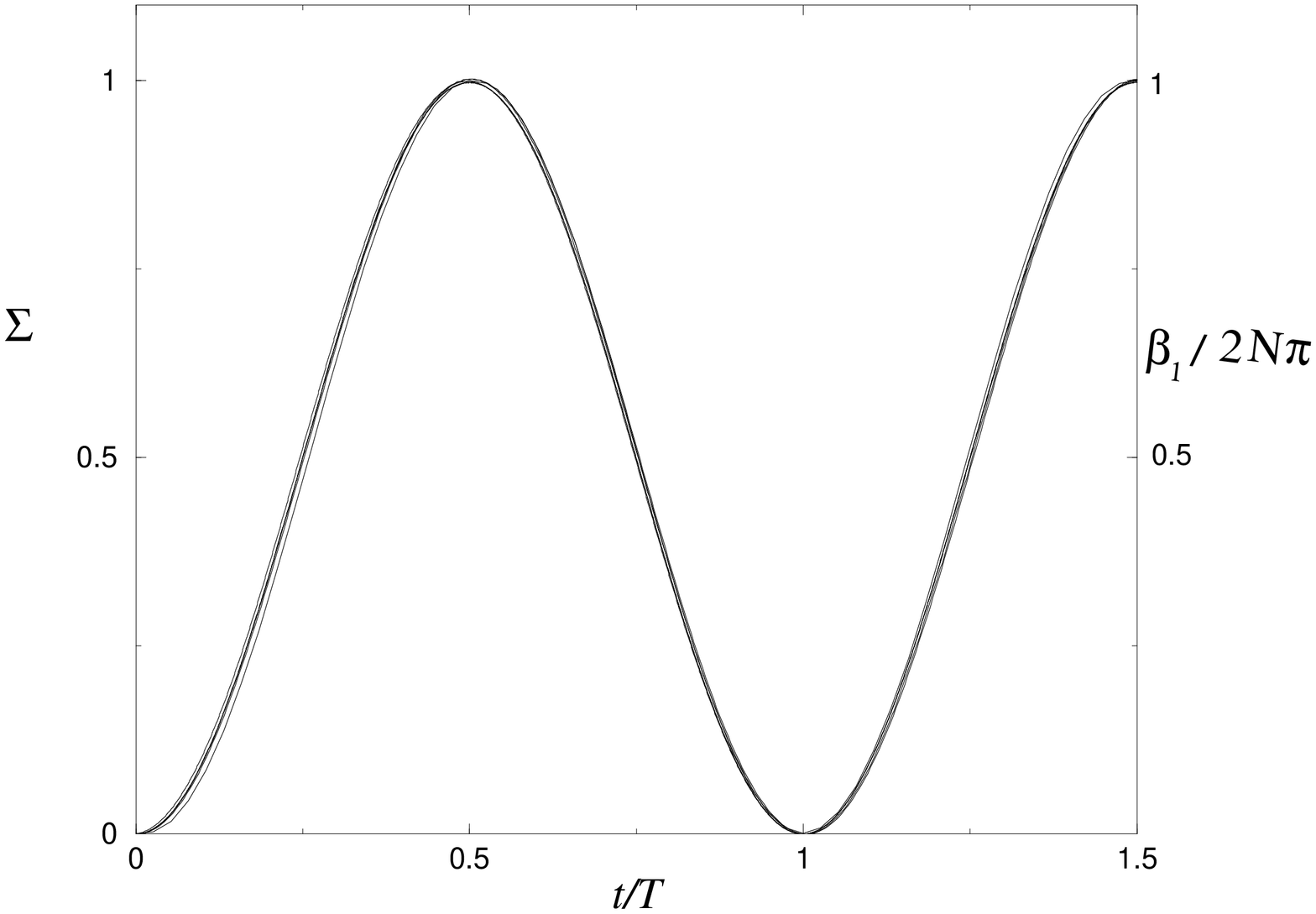,angle=270,width=12cm}
\caption{\small The scaled energy, $\Sigma \equiv (E\alpha 
^{-2}-E_k)/(E_n-E_k)$ where $E_k$ and $E_n$ are the unperturbed 
eigenenergies, 
and geometric phase acquired at each cycle, $\beta _1 / 2N\pi$,
vs.~$t/T$, $T$ being the oscillation period of the geometric phase, 
for the case of cylindrical cavity                   
at $\tilde{\omega}=66.632,\ 17.278$, and 22.21227,
using $\epsilon = 0.01$.}
\label{fig2}
\end{figure}

\begin{figure}
\psfig{file=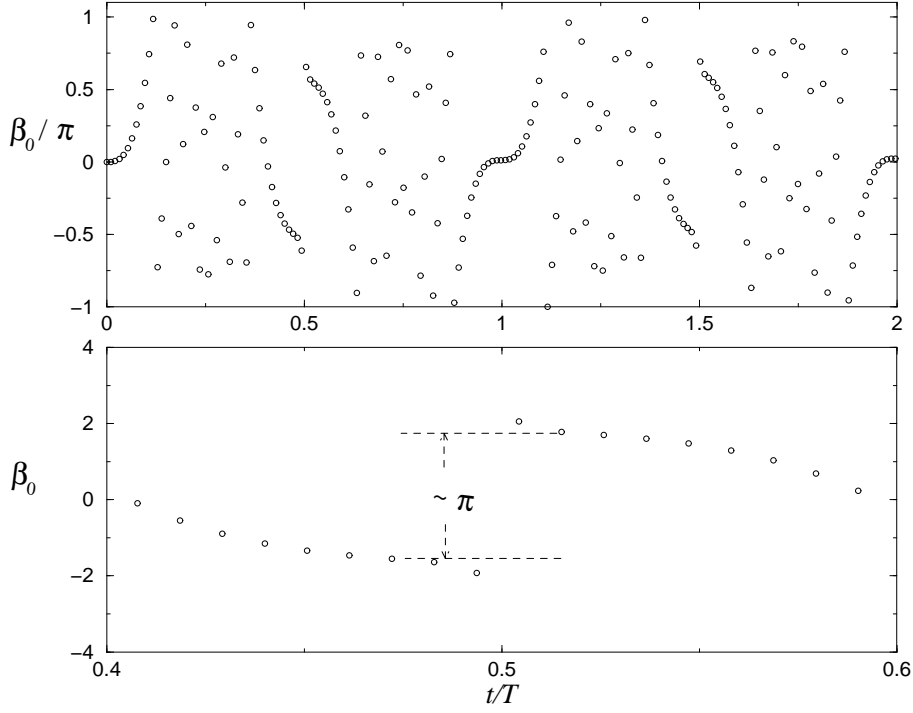,angle=270,width=12cm}
\caption{\small $\beta_0$ vs.~$t/T$ at $\tilde{\omega}=12.344$ for a
cylindrical cavity with $\epsilon=0.01$.  The region around $t/T = 0.5$ is 
enlarged and shown in the lower panel to show the approximate $\pi$-jump.
}
\label{fig3}
\end{figure}

\begin{figure}[htbp]
\psfig{file=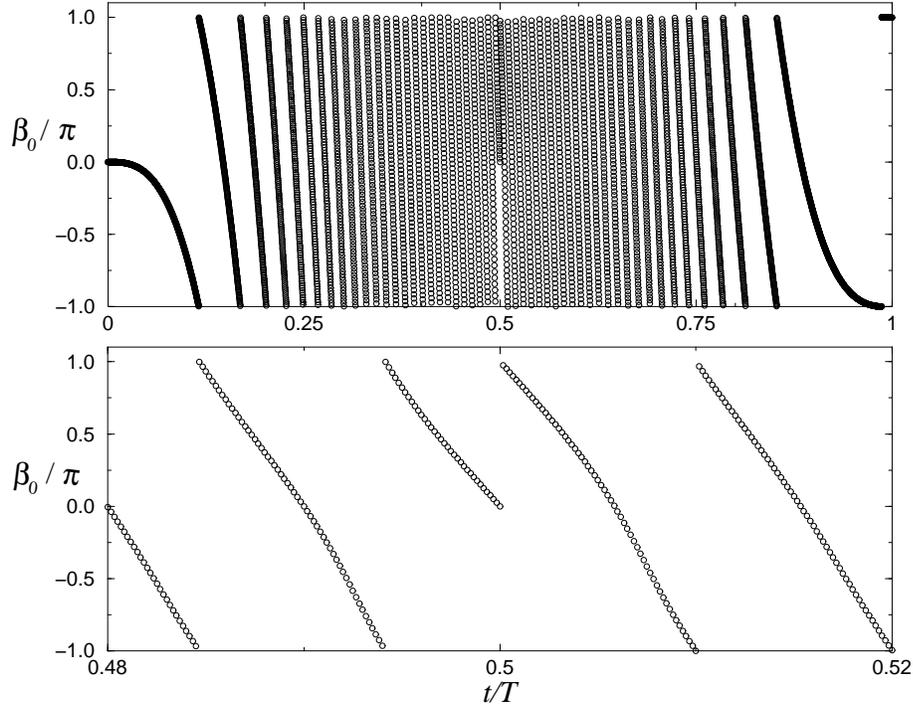,angle=270,width=12cm}
\caption{\small Same as Fig.~3, but for a spin-1/2 particle in a rotating 
magnetic field, $\alpha=1/100$, and for $t\neq n\tau$.}
\label{ebb0k100w60}
\end{figure}

\begin{figure}[htbp]
\psfig{file=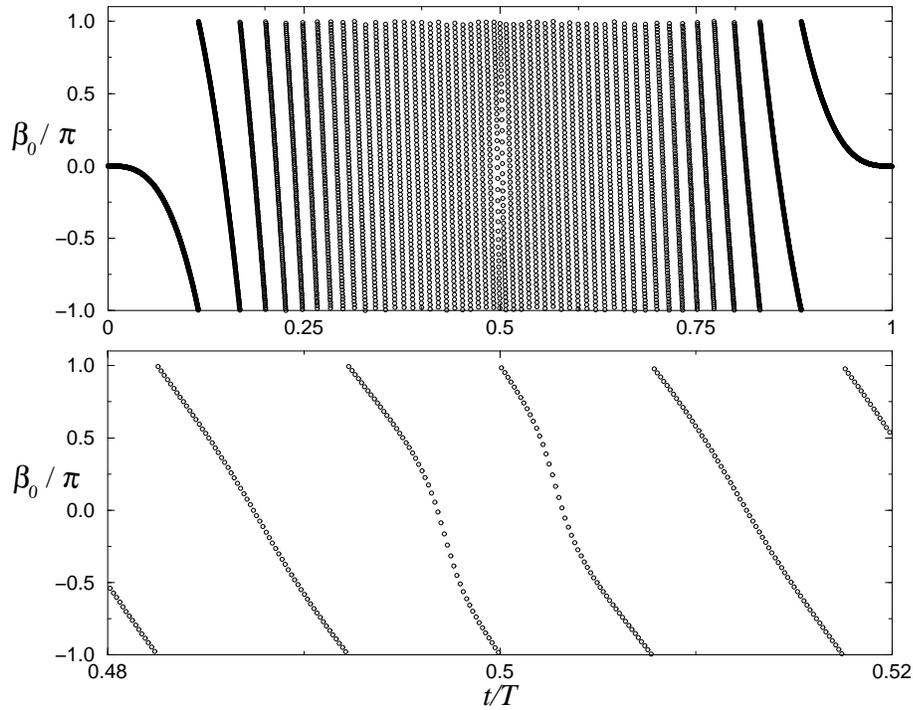,angle=270,width=12cm}
\caption{\small Same as Fig.~4, but for $\alpha = 1/101$.}
\label{ebb0k101w61}
\end{figure}

\begin{figure}[htbp]
\psfig{file=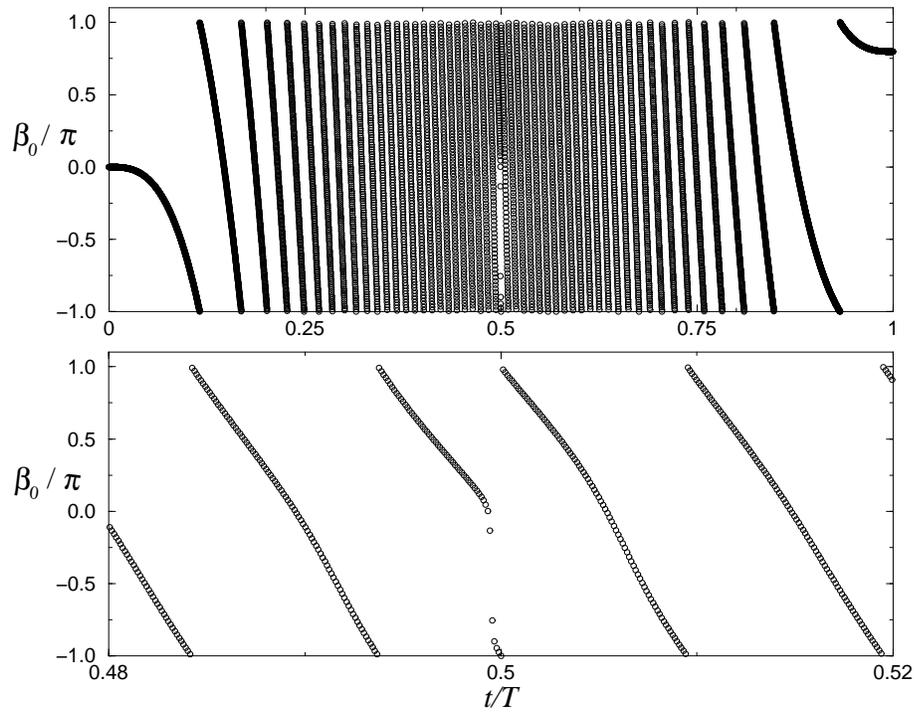,angle=270,width=12cm}
\caption{\small Same as Fig.~4, but for $\alpha=5/501$.}
\label{ebb0k100.2w90}
\end{figure}

\begin{table}[htbp]
\begin{center}
\vspace{4mm}
\begin{tabular}{|c|c||c|c|c|} \hline
  \multicolumn{2}{|c||}{Resonances} & 
  \multicolumn{2}{c|}{$\tilde{\Gamma}$} & $T\Gamma/\pi$   \\ \hline
N & n & Numerical & RWA   &    Numerical  \\ \hline
1 & 2 & 6.17 (7.45)   & 6.17 (7.40) &  1.00 (1.01) \\ \hline
1 & 3 & 16.6          & 17.3        &  0.98        \\ \hline
1 & 4 & 33.2          & 33.3        &  1.00        \\ \hline
\hline
2 & 2 & 4.15 (4.66)   & 4.63 (5.55) &  0.99 (0.96) \\ \hline
2 & 3 & 50.0 (76.5)   & 13.0 (14.8) &  1.00 (1.02) \\ \hline
2 & 4 & 891  (532)    & 25.0 (27.8) &  1.09 (1.17) \\ \hline
2 & 5 & 136           & 40.7        &  0.93        \\ \hline
\hline
3 & 4 & 205  (341)    & 23.6 (26.2) &  1.02 (1.07) \\ \hline
3 & 5 & 15700         & 38.5        &  1.02        \\ \hline
3 & 6 & 78300         & 56.8        &  1.07        \\ \hline
\end{tabular}
\caption{\small Scaled widths 
($\tilde{\Gamma} \equiv \mu R_0^2 \Gamma /\hbar\epsilon^{N}\eta_{nk}$)
of the lowest $k=1$ resonances for cylindrical cavity with $\epsilon = 0.01$.  
Full numerical results are compared to the RWA values.
The widths multiplied by the Rabi oscillation period, both extracted 
from the numerical data, are also shown in the last column.  Corresponding
numbers for spherical cavity with $\epsilon = 0.02$ are shown in parentheses.}
\end{center}
\end{table}


\begin{thebibliography}{33}

\bibitem{bag}P.~Hasenfratz and J.~Kuti, Phys.~Rep.~{\bf 40}, 75 (1978);
K.~Colanero and M.-C.~Chu, Phys.~Rev.~C {\bf 65}, 045203 (2001); 
K.~Colanero, Phd thesis, Chinese University, 2001. 

\bibitem{sono}B.P.~Barber {\it et al.}, Phys.~Rep.~{\bf 281}, 65 (1997).

\bibitem{qed}G.T.~Moore, J.~Math.~Phys.~{\bf 11}, 2679 (1970);
P.W.~Milonni, {\it The Quantum Vacuum} (Academic Press, New York, 
1993); N.D.~Birrell and P.C.W.~Davies, 
{\it Quantum Fields in Curved Space}  
(Cambridge University Press, Cambridge, 1982). 

\bibitem{bh}S.W.~Hawking, Nature {\bf 248}, 30 (1974); 
Commun.~Math.~Phys.~{\bf 43}, 199 (1975).


\bibitem{photon}For example, C.K.~Law, Phys.~Rev.~Lett.~{\bf 73}, 
1931 (1994); V.V.~Dodonov and A.B.~Klimov, Phys.~Rev.~A {\bf 53}, 
2664 (1996), and references therein; A.~Lambrecht, M.T.~Jaekel and 
S.~Reynaud, Phys.~Rev.~Lett. {\bf 77}, 615 (1996);
C.K.~Cole and W.C.~Schieve, Phys.~Rev.~A {\bf 52}, 
4405 (1995); 
P.~Meystre {\it et al.}, J.~Opt.~Soc.~Am.~B {\bf 2}, 1830 (1985);
Ying Wu {\it et al.},
Phys.~Rev.~A {\bf 59}, 3032 (1998);
Ying Wu {\it et al.}, Phys.~Rev.~A {\bf 59}, 
1662 (1998); 
K.W.~Chan, Master Thesis, The Chinese University of Hong Kong
(unpublished), 1999; K.~Colanero and M.-C.~Chu, Phys.~Rev.~E {\bf 62}, 
8663 (2000).


\bibitem{berry}M.V.~Berry, Proc.~Roy.~Soc.~Lond.~A {\bf 392}, 45 (1984).

\bibitem{Alfred}A.~Shapere and F.~Wilczek, {\it Geometric Phase in Physics} 
(World Scientific, Singapore, 1989).


\bibitem{Hilbert}Y.~Aharonov and J.~Anandan, Phys.~Rev.~Lett.~{\bf 58}, 
1593 (1987).

\bibitem{moore}D.J.~Moore, Phys.~Rep.~{\bf 210}, 1 (1991).

\bibitem{bird}D.M.~Bird and A.R.~Preston, Phys.~Rev.~Lett.~{\bf 61},
2863 (1988).

\bibitem{wagh}A.G.~Wagh and V.C.~Rakhecha, Phys.~Lett.~A {\bf 148}, 17 (1990)
.
\bibitem{klaus}K.~Colanero and M.-C.~Chu, Phys.~Rev.~A {\bf 60}, 1845 (1999).

\bibitem{gs}J.~Samuel and R.~Bhandari, Phys.~Rev.~Lett.~{\bf 
60}, 2339 (1988).

\bibitem{c.m.cheng}C.~M.~Cheng and P.~C.~W.~Fung, J.~Phys.~A: 
Math.~Gen.~{\bf 22}, 3493 (1989);
K.M.~Cheng and P.C.W.~Fung, J.~Phys.~A: 
Math.~Gen.~{\bf 23}, 4851 (1990). 

\bibitem{sakurai} J.J.~Sakurai, {\it Modern Quantum Mechanics} 
(Addison-Wesley, Redwood City, 1985).

\bibitem{griffiths} David J. Griffiths, {\it Introduction to Quantum Mechanics} 
(Prentice Hall, Upper Saddle River, 1995).

\bibitem{R. Bhandari} R.~Bhandari, Phys.~Rev.~Lett.~{\bf
88}, 10 (2002).

\end{thebibliography}
\end{document}